# Structural Order of the Molecular Adlayer Impacts the Stability of Nanoparticle-on-Mirror Plasmonic Cavities

Aqeel Ahmed,[†] Karla Banjac,[‡] Sachin S. Verlekar,[†] Fernando P. Cometto,[‡,¶]

Magalí Lingenfelder,[*,‡] and Christophe Galland[*,†]

[†]Laboratory of Quantum and Nano-Optics and Institute of Physics, École Polytechnique Fédérale de Lausanne, CH-1015 Lausanne, Switzerland.

[‡]Max Planck-EPFL Laboratory for Molecular Nanoscience and Institute of Physics, École Polytechnique Fédérale de Lausanne, CH-1015 Lausanne, Switzerland.

[¶]Departamento de Fisicoquímica, Instituto de Investigaciones en Fisicoquímica de Córdoba, INFIQC–CONICET, Facultad de Ciencias Químicas, Universidad Nacional de Córdoba, Ciudad Universitaria, X5000HUA Córdoba, Argentina.

E-mail: magali.lingenfelder@epfl.ch; chris.galland@epfl.ch

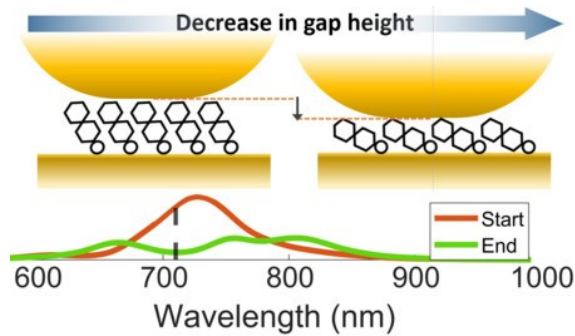

## Abstract

Immense field enhancement and nanoscale confinement of light are possible within nanoparticle-on-mirror (NPoM) plasmonic resonators, which enable novel optically-activated physical and chemical phenomena, and render these nanocavities greatly sensitive to minute structural changes, down to the atomic scale. Although a few of these structural parameters, primarily linked to the nanoparticle and the mirror morphology, have been identified, the impact of molecular assembly and organization of the spacer layer between them has often been left uncharacterized. Here, we experimentally investigate how the complex and reconfigurable nature of a thiol-based self-assembled monolayer (SAM) adsorbed on the mirror surface impacts the optical properties of the NPoMs. We fabricate NPoMs with distinct molecular organizations by controlling the incubation time of the mirror in the thiol solution. Afterwards, we investigate the structural changes that occur under laser irradiation by tracking the bonding dipole plasmon mode, while also monitoring Stokes and anti-Stokes Raman scattering from the molecules as a probe of their integrity. First, we find an effective decrease in the SAM height as the laser power increases, compatible with an irreversible change of molecule orientation caused by heating. Second, we observe that the nanocavities prepared with a densely packed and more ordered monolayer of molecules are more prone to changes in their resonance compared to samples with sparser and more disordered SAMs. Our measurements indicate that molecular orientation and packing on the mirror surface play a key role in determining the stability of NPoM structures and hence highlight the under-recognized significance of SAM characterization in the development of NPoM-based applications.

# Introduction

Metallic nanostructures are known to support localized surface plasmon resonances (LSPRs) in the visible and near-infrared range. When placed in close proximity, the LSPRs supported by individual nanoparticles hybridize to form new modes,[1] some of them providing a large enhancement of the incident electromagnetic (EM) field inside the gap or dielectric layer between the nanostructures.[2] These new LSPRs and the corresponding magnitude of electromagnetic field enhancement depend on a number of factors including the material composition, morphology, and dimensions of the nanostructures and the spacer between them.[3] Several hundred-fold field enhancement along with sub-nanometric spatial sensitivity of such structures are utilized in numerous applications: sensing and chemical fingerprinting with surface enhanced Raman scattering (SERS),[4,5] nonlinear generation of light,[6,7] strong light matter interaction,[8,9] nanoscale chemistry,[10,11] photodetection,[12] and integrated electrical-to-optical conversion,[13,14] to cite a few examples. Plasmonic nanocavities can also function as nanoscale sources of heat due to the Joule effect[15] and can be employed for several biological and medical applications.[16,17] Owing to the above mentioned and many more potential applications, numerous plasmonic nanostructures have been developed over the years.[18,19] While state-of-the-art photolithographic techniques can be used for structures with critical dimensions between 10 nm – 100 nm, reproducible fabrication of sub-10 nm devices is best achieved using molecular self-assembly, which enables the realization of nanocavities with sub-nanometer gap sizes.[20,21]

Among these nanostructures, the nanoparticle-on-mirror (NPoM) geometry offers unpar-alleled ease of fabrication with substantial reproducibility. The NPoM consists of a nanoparticle positioned above a flat metal film (the mirror) with a well-defined gap between them. When the NPoM is illuminated, the charge oscillations initiated in the nanoparticle form an inverted image within the mirror surface; their mutual coupling results in the overall plasmon resonances. The fundamental properties of these plasmonic modes have been investigated in a number of pioneering works.[22-25] Commonly found plasmon modes in the NPoM are the transverse mode of the nanoparticle and the bonding dipole mode (labeled here $l_i$) between the nanoparticle and the mirror. Moreover, higher order dipole modes and cavity modes may also exist in the gap due to facets present on the surface of the nanoparticle.[26]

To the best of our knowledge, the first demonstration of NPoM can be traced back to the enhancement of light emission from metal-insulator-metal tunnel junctions that was first observed by McCarthy et al.[27] in 1977. A few groundbreaking experimental studies include the works of Holland et al.;[28] with silver flakes on top of silver films with lithium fluoride spacer, and Kume et al.;[29] with silver nanoparticles embedded in silica matrix on top of silver film. NPoM structures with SAM as spacer layer were introduced at the beginning of the 21[st] century by Hutter et al.[30] and Okamoto et al.[31] The ease of fabrication and the reproducibility of SAM-based NPoM bypasses the limits imposed by conventional fabrication processes, and this approach has been shown to be compatible with deterministic positioning of the nanoparticles.[32] As the gap size is decreased below 1 or 2 nm, quantum mechanical effects can be observed at room temperature.[33] The immense EM field enhancement and extremely small mode volume make the LSPRs of these nanocavities susceptible to minute changes in the NPoM structure,[3] especially to the structure and configuration of SAM layers.[34] Conformational changes of molecules within a break junction are known to affect the conductivity of the junction.[35,36] While the adsorption of thiolated molecules on noble metals has been extensively studied,[37,38] the dynamics of molecular species within the SAM and its impact on the optical response of NPoM remain unclear.

Decades of fundamental surface science studies combining local spectroscopies, electro-chemistry, and theoretical modelling[39,40] have shed light into the two-step process of the SAM formation.[41,42] The first step is characterized by fast adsorption and desorption processes, where thiol molecules adsorb to the Au surface in lying-down conformation (i.e. with the principal molecular axis being near-parallel to Au surface) and remain highly mobile.[41] The second step leads to slow re-organization of the initial phase into a densely packed phase in which the molecules stand in the upright configuration (with the principle molecular axis being close to the surface normal).[43] SAM re-organization partially proceeds through the transport of the thiol-Au ad-atom moieties that remain mobile once the strong thiol-Au bond is formed.[44] This process leads to the formation of Au islands[45] – a process known as surface reconstruction.

The two-step process generally describes the formation of thiol-based SAMs on Au. However, the adsorption kinetics and its final SAM structure strongly depend on the surface conditions of the Au substrate and on the preparation protocol. The critical surface conditions affecting SAM ordering are: the size of the crystalline facets, the orientation of the surface facets (expressed as ratio of [111]:[100]:[110] facets),[46] and the presence of atomic and nanometer-sized (e.g. grain boundaries) surface defects. The main parameters of the preparation protocol affecting SAM ordering are: the solvent,[47] the temperature,[48] and the incubation time. The complexity of the SAM configuration expressing multiple domains on the Au surface has been observed even in densely packed SAMs prepared on monocrystalline surfaces. Local studies using scanning tunneling microscopy (STM) reveal significant heterogeneities in SAM ordering, even on spatial scales below 50 nm. Having in mind the parameters affecting the SAM preparation, it becomes clear that perfect single domain SAM layers are a rare exception rather than reality, especially on polycrystalline Au substrates.

In this article, we study how the organization and orientation of biphenyl-1,4-thiol (BPhT) molecules within NPoM plasmonic nanocavities affect their optical properties, and how these properties are modified under focused near-infrared laser excitation with power levels ranging from few $\mu W/\mu m^2$ to $mW/\mu m^2$. The molecular orientation and packing den-sity is tuned using variable incubation time of the gold film in the BPhT solution. For clarity we limit ourselves to two incubation times, 2 hours (labeled S2) and 24 hours (labeled S24), ensuring significantly different SAM morphologies. The differences between S2 and S24 were assessed using reductive desorption and STM, confirming that the longer incubation of S24 leads to denser packing of the molecules and more pronounced surface reconstruction of the gold film compared to the shorter incubation of S2, which yields a sparser monolayer of lying-down molecules. Many individual NPoMs on each type of sample were optically characterised using darkfield scattering and Raman spectroscopy under increasing laser illumination power. We find that changing the SAM incubation time has a significant impact on the plasmon resonance wavelength and on the stability of the NPoM cavities, with the observation that a more ordered and densely packed monolayer results in a less stable plasmonic response, rapidly red-shifting and degrading under laser illumination. We attribute this observation to heat-induced reorientation of the molecules toward a flatter configuration, together with the increased mobility of gold adatoms after surface restructuring, which facilitate the formation of conductive bridges between the gold film and the nanoparticle in samples with longer incubation times. Our study reveals the overlooked critical role of SAM preparation conditions in fabricating robust and reproducible NPoM plasmonic resonators. It contributes to a better understanding of plasmonic nanojunctions and should stimulate further research at the interface between plasmonics and surface science in order to push nano-plasmonics closer to applications.

## Results and Discussion

### Electrochemical (EC) Characterization

We characterize the molecular ordering of BPhT SAMs using reductive desorption ($Au-S-C_6H_4-C_6H_5 + e^- \longrightarrow Au_0 + S-C_6H_4-C_6H_5$) with the specific signature being cathodic waves. The peak positions, i.e. potential of the reductive desorption, strongly depend on the lateral inter-actions between BPhT molecules[46,49] and are indirectly related to the density of the SAM packing. The desorption peak of densely packed SAMs is thus shifted to the more negative potentials (Table S1). The peak width and the area under the peak depend on the number of the thiol molecules desorbing from the Au surface at a given potential. Broad desorption peaks are thus related to the ill-defined SAMs of low-packing density. We note that EC characterization is a global characterization technique where SAM desorbs over the whole sample surface. It provides information on the molecular ordering over the macroscopic scale.

Figure 1a shows the cyclic voltammograms of S24 and S2 prepared on glass/Cr/Au substrate. Cathodic peaks at -0.88 V and -0.85 V versus Ag/AgCl reference electrode correspond to reductive SAM desorption on S24 and S2, respectively. Since desorption of highly-ordered SAM proceeds at more negative potentials, the shift of $\approx$ -30 mV for S24 suggests a high degree of SAM ordering. Moreover, broad desorption peak on S2, with the center slightly shifted towards positive potentials, implies that BPhT SAM on S2 is globally in ill-defined phase. We also note the fine structure in the S2 voltammetric wave (Figure 1a inset). The shoulder shifted to the negative potentials with respect to the main peak might arise from small SAM domains (< 50 nm$^2$) being in the phase of denser packing than ill-defined phase characterized by the main S2 desorption peak (Fig. S5).

## Scanning Tunneling Microscopy (STM)

SAMs formed upon two different incubation times, i.e. S2 and S24, were also studied using STM. We highlight that STM characterization was performed on the mirror substrate (i.e. template stripped Au) following the same procedure as for NPoM preparation. As discussed earlier, SAM characterization on the mirror substrate is of great importance since molecular ordering depends on surface crystallinity, the size of the terraces, and the presence of surface defects.

Figure 1b and 1c show STM images of S24 and S2 over 45 $\times$ 45 nm$^2$ Au surface functionalized with SAMs: there are clear differences both in the morphologies of Au surface and in SAM ordering. Large scale STM images of S24 reveal the presence of Au adatom islands and pits formed upon lifting Au reconstruction (Fig. S2). High-resolution STM images of SAM layer on S24 show regular patterns characteristic for the densely-packed phase (Fig. S3 and Fig. S4) where the thiol molecules are in up-right conformation[50] as illustrated on the scheme in Figure 1b. This assignment agrees with UHV-STM studies reporting densely packed BPhT SAMs in up-right conformation prepared upon long incubation.[51,52] Additional STM images (Fig. S1) reveal the co-existence of at least two different densely-packed SAM phases with the molecules in the up-right conformation on neighboring Au facets. As we shall see, the coexistence of different phases leads to the angstrom level deviations in the gap size heights measured for different NPoMs, even along the same sample (Fig. S1).

STM images of BPhT SAM on S2 show corrugations in the angstrom range without a well-defined pattern observed on S24 (Figure 1c). Large scale STM images appear fuzzy: while the SAM layer fully covers the surface, the fuzziness is related to the mobility of BPhT molecules. The absence of a periodic pattern suggests that the SAM on S2 is less-densely packed than the SAM on S24. Both sparser packing and higher mobility of the molecules suggest that the SAM on S2 are mostly in the lying-down phase, with the principal molecular axis making a large angle with respect to the surface normal. To the best of our knowledge, this is the first STM observation of the lying-down phase of BPhT SAM at the liquid/solid interface. Moreover, we observe small SAM domains (Fig. S5) having a high packing density in agreement with the fine structure of S2 desorption peak (Figure 1a inset).

## Optical Characterization

The spectral position of the dipole mode is known to be extremely sensitive to structural parameters of the NPoM such as nanoparticle size and shape along with the height, refractive index, and conductivity of the gap layer.[53-55] We study about 10 NPoMs on each sample. Figures 1(f,g) show the DF spectra measured before any laser exposure on the two samples. The initial position of the dipole mode recorded on S24 (Figure 1f) is blue-shifted compared to S2 (Figure 1g). This relative shift can be attributed to the difference in gap height due to distinct molecular orientations. In the long incubation time sample (S24), the molecules adopt an upright orientation: hence the gap height is approximately equal to the length of the molecule – we use 1.4 nm in the simulation shown as dashed line in Figure 1f. With a shorter incubation time (S2), the molecules adopt the lying-down configuration leading to a smaller gap – we find good agreement with 0.8 nm gap height in the simulation shown as dashed line in Figure 1g. Note that in both cases the molecules within the gap have a significantly large electrical impedance, as expected for BPhT – a more conductive gap would result in a blue shift of the dipole mode by as much as 50 nm.[54] These observations are in perfect concordance with the STM characterisation discussed above. To our knowledge, it is the first time that a clear correlation between incubation time, SAM morphology and plasmonic response of the NPoMs is established using a combination of EC, STM, and optical techniques.

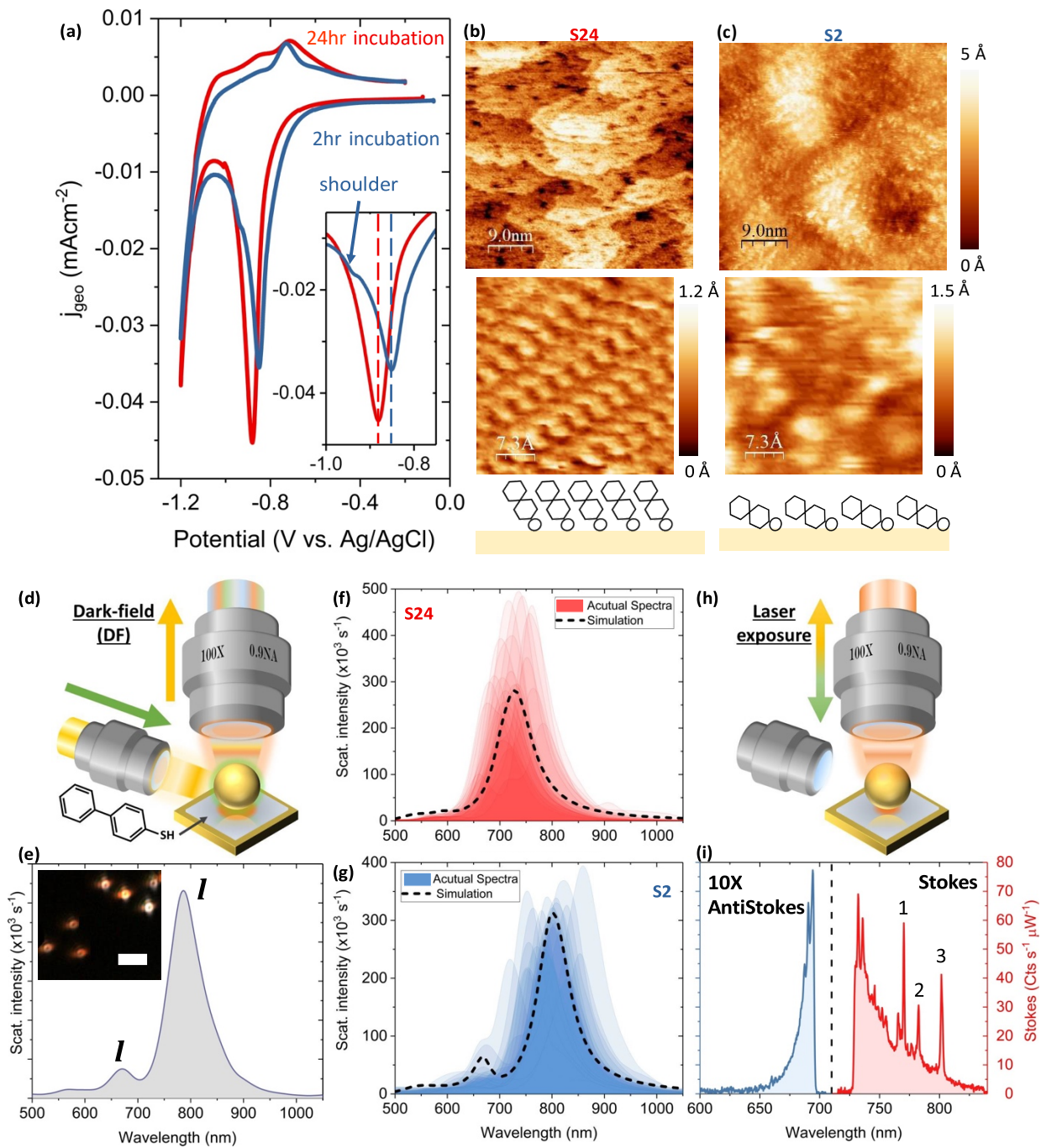

Figure 1: (a) EC characterization of the BPhT SAM: cyclic voltammograms showing the reductive desorption (oxidative adsorption) in the negative (positive) direction of BPhT SAM prepared upon incubation of glass/Cr/Au substrates during 24 hours (S24, red solid line) or 2 hours (S2, blue solid lines). Inset shows an enlarged view of the desorption SAM peaks. (b) STM images of BPhT SAM on S24. The large scale STM image (upper panel) shows the terraces and the pits formed upon Au surface reconstruction. Small scale STM image (middle panel) shows the pattern characteristic for the densely packed BPhT SAM phase illustrated in the scheme (lower panel). (c) STM images of BPhT SAM on S2. Large scale STM image (upper panel) shows a corrugated Au surface with randomly disturbed dot-like features. Small scale STM image (middle panel) shows the dot-like features appearing fuzzy. Scanning parameters for bias voltage ($V_{bias}$) and tunneling current ($I$): (b) $V_{bias}$= 200 mV, $I$ =338.2 pA and $V_{bias}$=-336 mV, $I$ =420 pA, (c) $V_{bias}$=586 mV, $I$ =567.6 pA and $V_{bias}$=532 mV, $I$ =482.1 pA. (d) Schematics for dark-field (DF) scattering spectroscopy. White light is used to illuminate the NPoM from the side (green arrow), while the scattered light is collected (yellow arrow) via a 0.9 NA objective. (e) Example of DF spectrum showing bonding dipole plasmon $l_1$ mode and a higher order dipole mode $l_2$. The inset is a DF image of the sample showing multiple NPoMs. Scale bar is 1 $\mu$m. DF spectra from all NPoMs measured across (f) S24 and (g) S2. All the spectra are smoothed by moving average of 20 points. The black dashed lines indicate the simulated scattering spectra obtained with a gap height of 1.4 nm in (f) and 0.8 nm in (g); for both the refractive index of the gap and AuNS facet size were 1.4 and 30 nm, respectively. They were computed by boundary element method (BEM) using MNPBEM MatLab package[56] and arbitrarily rescaled in amplitude to fit in the figure. (h) During laser exposure, light is focused on the NPoM and the Raman signal collected through the same objective. (i) Example of Raman spectrum. The dashed line represents the laser wavelength at 710 nm. The Raman modes of BPhT are clearly visible on the Stokes (red line) side where 1, 2, and 3 symbolize the main Raman peaks for BPhT at 1079, 1281 and 1586 $cm^{-1}$ respectively. For the anti-Stokes (blue color), the electronic Raman background is dominant, and can be used to estimate the electronic temperature.

After the measurement of the initial DF spectrum, each NPoM was illuminated with a tightly focused laser beam tuned at 710 nm (wavelength in vacuum) (Figure 1h) and the Raman spectrum (Figure 1g) from BPhT was measured after filtering the laser line, in order to probe the integrity of the molecules. The laser power was first set to the lowest value of $\sim$ 8 $\mu$W and the NPoM was exposed to the laser for 5 minutes, during which the Raman signal from BPhT is recorded. The camera exposure time per frame and the number of frames per measurement were adapted depending on the strength of the Raman signal, while keeping the total laser exposure to 5 minutes for each power. After laser exposure the DF spectrum is measured, and the cycle is repeated after an increase of the laser power.

Examples of typical changes to the DF spectra upon laser exposure for the two different samples are shown in Figure 2(a) and 2(b). For S24 the dipole mode appears to redshift already after first laser exposure with $\sim$ 8 $\mu$W, without significant change in the peak scattering intensity. The DF spectra measured after each exposure show that the dipole mode continues to red shift along with declining peak intensity as the laser is ramped to higher powers. After the fifth exposure with about 170 $\mu$W, the dipole mode has shifted close to 800 nm and the intensity has dropped noticeably compared to the initial value (Figure 2a, green line). The higher order dipole mode has also red shifted to 700 nm with an increase in peak intensity. Upon further increase of laser power the spectrum suddenly blue-shifts, indicating the formation of a conductive bridge between the AuNS and the mirror underneath and finally the AuNS is completely fused with the mirror (Fig. S8).[57]

The evolution is quite different for the NPoMs on S2, which is the sample with disordered SAM with lying-down molecules. In the example shown in Figure 2(b), the dipole mode is initially located around 800 nm (blue line in lower panel) and does not appear to red shift as readily as observed for S24 when the laser power is increased. The peak intensity also remains unchanged as measured after each exposure. It is only after laser powers greater than 400 $\mu$W that a significant red-shift and a decrease in intensity of the dipole mode are observed. Bridge formation followed by fusion of AuNS to the mirror similar to that observed for S24 occurs at even higher powers (Fig. S9).

Multiple NPoMs across S24 and S2 studied with the same procedure disclose the distinct and layer dependent evolution of the dipole mode with increasing laser power, as presented in Figure 2(c,d). The various red (blue) polygons indicate individual NPoMs on S24 (S2) and the red (blue) shaded region shows the spread of the standard deviation centered at the mean. Differences in the SAM ordering across the substrate result in inhomogeneous layer height and contribute to the spread in resonance position, together with the distribution of shape and size of the nanoparticles. As the laser power is progressively increased, the NPoMs on S24 show a prompt red-shift of the dipole mode while the NPoMs on S2 are stable up to powers above 300 $\mu$W, after which they display a significant red-shift too. In term of intensity of the dipole mode, the S24 NPoMs show a quick drop while S2 NPoMs remain stable. With this study, we evidence the fact that NPoM cavities formed on top of a disordered and loosely packed SAM, obtained with short incubation time (S2), retain their pristine optical and plasmonic properties for laser powers that are an order of magnitude larger than what NPoMs formed on a more ordered and densely packed monolayer (S24) can sustain. This observation is a central result, which was not anticipated and should become a precious guide for future optimization of plasmonic nanocavities in view of their diverse applications.

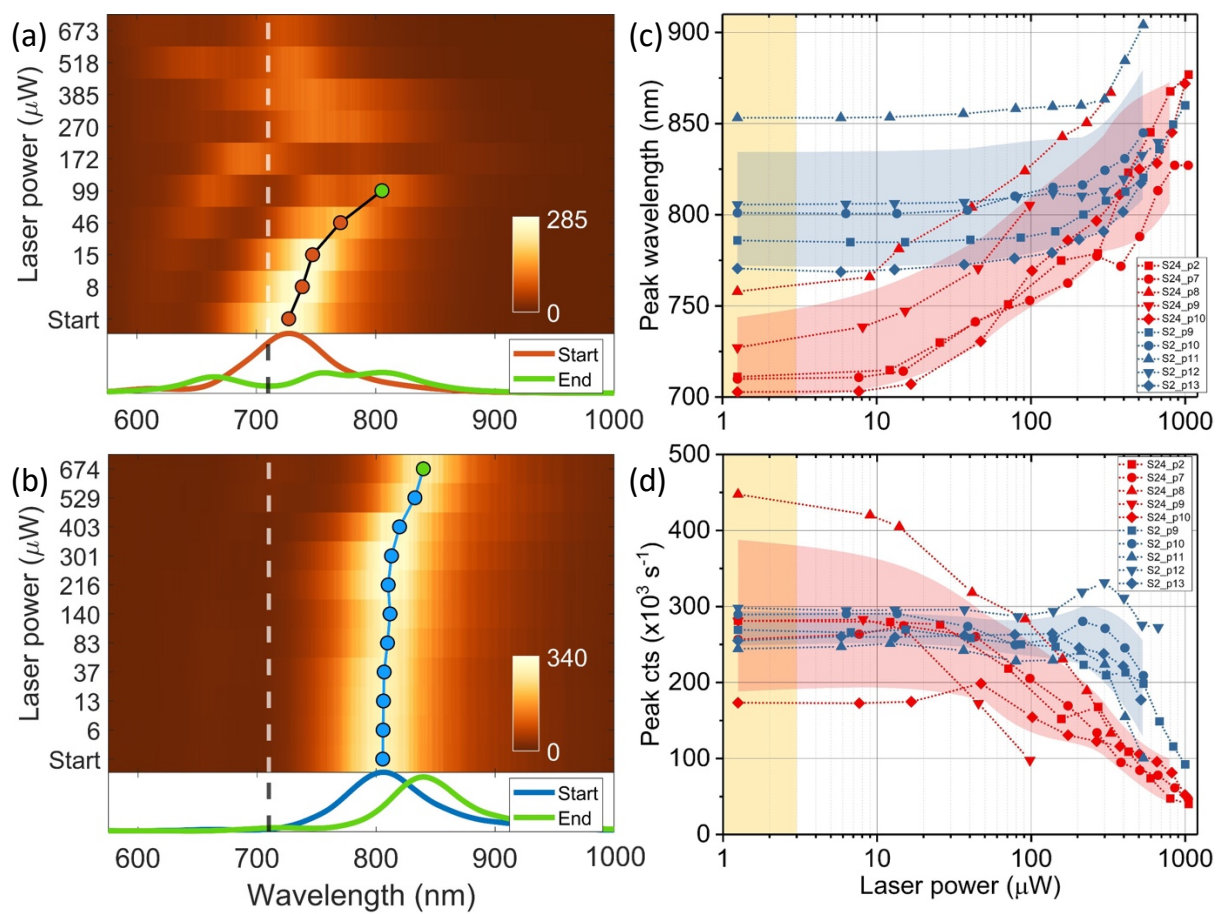

Figure 2: (a, b) Color plots showing the evolution of DF spectra upon repeated laser exposure with increasing power for samples incubated for (a) 24 hours (S24) and (b) 2 hours (S2). The dots and line show the evolution of the dipole mode in each case, the color scale represent the counts per second recorded on the spectrometer, and the dashed white line represents the laser wavelength used for exposure. The panel underneath each color plot shows the initial (red for S24 and blue for S2) and final spectra (green). (c) The peak scattering wavelength and (d) intensity of the dipole mode of multiple NPoMs across two samples with different SAMs. The various NPoMs measured on S24 and S2 are denoted by red and blue symbols respectively. The shaded region represents the extent of standard deviation around the mean value over all NPoMs of each sample. The individual DF spectra for each measured NPoM are shown in Fig. S8 and Fig. S9.

We now turn our attention to the understanding of the physical transformations underlying the changes in scattering spectra. A number of different phenomena may occur during the laser exposure of the NPoM structure. Here, limiting the study to sub-mW laser powers, we consider two physical mechanism likely responsible for the evolution of the scattering spectra: increase in facet size (Figure 3a) and decrease in gap height (Figure 3c). We use numerical simulations to predict their effects on DF spectra and consider these two processes independently; e.g. the facet size is gradually varied while keeping the gap size constant, and vice versa. The initial facet size of commercially available AuNS used for this study (80 nm diameter, BBI Solutions) is approximately 30 nm,[53] and it was suggested that in similar systems the facet size can be irreversibly increased by sweeping the power of an incident laser.[57] To simulate the scattering spectra with different facet sizes (Figure 3b) the NPoM was modelled by placing an initially faceted AuNS above a 1.4 nm thick dielectric layer with a refractive index of 1.4. Starting from a facet diameter of 20 nm the dipole mode slightly red shifts until the facet diameter is close to 30 nm and then begins to blue shift. While the behavior of the dipole mode has been explained before,[58] it clearly does not match our results. In contrast, our results match well with the scenario where the gap height gradually decreases (Figure 3c). Simulating the scattering spectra of NPoM with gap height decreasing from 1.4 nm to 0.5 nm; and with constant facet size and refractive index of 30 nm and 1.4 respectively, a red shift of the dipole mode is observed, similar to our observations. In order to quantitatively compare simulations and experimental results the average initial (label i) and final (label f) positions of the dipole mode as observed on S2 (blue) and S24 (red) are plotted on top of the simulated data. The error bar represents the respective standard deviation. We conclude that, in contrast with previous reports invoking mostly the growth of facet size,[57,59] the shrinkage of the spacer layer and effective gap size is most consistent with our experimental observations and better explains the change in DF scattering spectrum caused by laser irradiation.

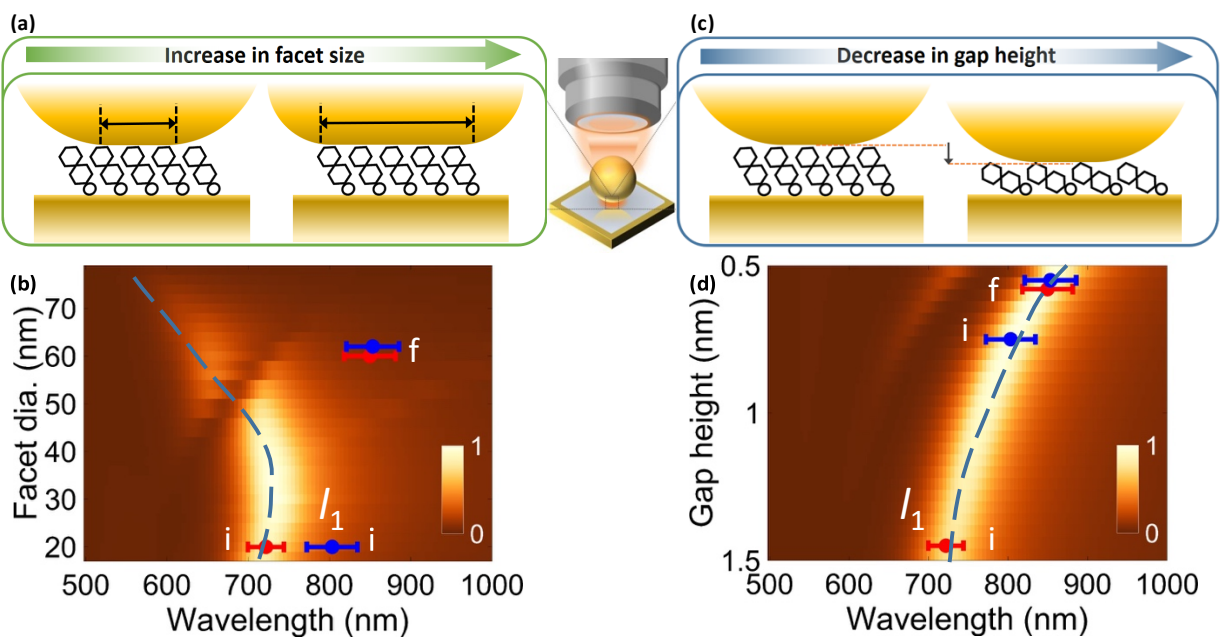

Figure 3: Two plausible explanations for the changes of the DF spectra. (a) The facet of AuNS may expand due to atomic migration under laser exposure; (b) BEM simulations show the evolution of the dipole mode $l_1$ as the facet diameter is gradually increased with constant gap height of 1.4 nm. (c) Alternatively, the gap height may decrease due to molecular reconfiguration; (d) corresponding simulation where the gap size is swept from 0.5 nm to 1.5 nm keeping the facet size fixed at 30 nm. In both simulations the molecules were modelled as a dielectric layer with refractive index of 1.4. The experimental values for the average initial (label i) and final (label f) wavelength of the dipole mode on S24 (red symbol) and S2 (blue symbol) are shown, with their positions along the $y$ axis adapted to match the simulations (note that it is not possible to reach a match in (b)). The error bars represent the standard deviation across multiple NPoMs.

The decrease in the gap height can be explained by thermally driven reorientation of the spacer molecules during laser exposure. It has been observed before that upright and densely packed SAM layers, similar to those on S24, may develop defects and switch to a lying down configuration as the temperature is increased to 400 K.[38,47,50,60] Temperatures ranging from 300 K up to a few thousand kelvin have been recorded in plasmonic nanogaps under optical excitation,[61] hence making thermally-activated molecular reorientation a plausible mechanism. In case where the SAM is not densely packed, similar to S2, the molecules are already in a lying down phase[51] and hence show little changes at low laser powers. Our hypothesis implies that, as the SAM height of the NPoMs on S24 decreases it should at some point match the SAM height of NPoMs on S2, and both samples should feature similar resonance positions. We find that this statement is in good agreement with the experimental results shown in Figure 2c. Importantly, though, measurements with laser wavelength at 632, 710 and 760 nm indicate that laser detuning from the the dipole mode does not have a major impact on the behavior of the NPoMs (Fig. S7), so that the difference in stability observed between S2 and S24 cannot be simply attributed to their different initial resonance position.

The changes in peak intensity of the dipole mode with laser power are slightly different for the two types of NPoMs (Figure 2d). While the continuous decrease in the peak intensity for NPoMs on S24 is correlated with the red-shift of the peak wavelength, the decrease in gap height alone cannot explain the decrease in peak intensity as shown by BEM simulations in Figure 3(d). This discrepancy is attributed to phenomena such as electron tunneling and charge screening that begin to play a significant role in sub-nanometer gap sizes.[33] Moreover, changes in electrical conductivity and electron tunneling abilities of molecules occur as their orientation is varied.[62–65] The laser-induced re-organization of the BPhT SAM in S24 from densely-packed phase to sparsely-packed phase with the lying-down molecules is very likely associated with the changes in conductivity and charge tunneling properties of molecules. The difference in the density of molecules within the NPoMs on S2 and S24 could also have an impact on the tunneling properties and the overall behavior of the dipole mode. Lastly, increase in AuNS facet size, Au adatom density, and restructuring of the Au surface during incubation are also important factors and may play a role in explaining the experimental results in all their details. Further work is required to develop a comprehensive understanding of metal-molecule plasmonic nanojunctions and how to tailor their stability and other properties by surface science and molecular engineering.

# Conclusion

In conclusion, by performing laser exposure with systematically increasing intensities and acquiring scattering spectra after each exposure, the changes within the nanogap of plasmonic cavities are tracked by analyzing the evolution of the dipole mode. Our results indicate that while laser exposure on the NPoM cavity may lead to restructuring of the Au nanoparticle, it cannot explain the dependence on incubation time of the dipole mode shift. For long incubation time (S24), the STM images as well as the initial position of the dipole mode indicate densely packed domains of upward standing molecules. As the laser power is increased the dipole mode red-shifts and decreases in scattering intensity. Both of these effects hint towards a decrease in the gap height due to thermally driven molecular reorientation. However, if the molecules are already in a lying down position, as shown by STM and DF for low incubation time sample (S2), the effect of laser exposure is less severe than that observed for long incubation. These observations stress the sensitivity of the NPoM systems based on molecular spacer layers to the details of their preparation. Our results emphasize that molecular parameters such as orientation and packing, as well as their evolution under the conditions prevalent within plasmonic nanocavities, must be studied and controlled in order enable the development of mature molecular and plasmonic optoelectronic devices.

# Methods

## Sample preparation

The NPoM nanocavities are fabricated starting with the mirror: an ultraflat Au film obtained by template stripping. [66] In order to facilitate the identification of individual NPoMs distinct markers are produced on the mirror surface by patterning the template. [67] The template is fabricated using 100 mm diameter <100> silicon (Si) test grade wafers with 100 nm silicon nitride ($Si_3N_4$) top layer. A 600 nm thick layer of positive photoresist AZ ECI3000 is spin-coated on the wafer. After a contact exposure and development, the nitride layer underneath the markers is etched using $CHF_3/SF_6$ in a reactive ion etcher until the Si surface is reached. The photoresist is then stripped and the wafer is placed in 40% potassium hydroxide (KOH) solution at 70°C for approximately 30 min. Finally, the $Si_3N_4$ layer is removed using hydrofluoric acid (HF) and the wafer is rinsed thoroughly with DI water and dried under nitrogen. The anisotropic etching of Si by KOH allows fabrication of markers without compromising the roughness of Si. As the etching only takes place along the crystallographic planes of Si, the etched side walls are significantly less rough than those obtained with other etching methods. A 200 nm thick Au film is deposited by electron-beam-evaporation on the Si template at room temperature with a rate of 5 Å/s and without any adhesion layer. For template stripping, glass chips are glued on the gold surface via epoxy (NOA61) and cured under a UV lamp for 45 minutes. In the absence of adhesion layer between Si and Au, the Au thin film is attached more firmly to the epoxy and is cleanly removed from the substrate when the chip is peeled off. These ultraflat Au substrates are then incubated at room temperature (approx. 22°C) in 1 mM solution of biphenyl-1,4-thiol (BPhT)(Sigma Aldrich, >99.5%) in ethanol. Two different incubation times are used: 2 hours (S2) and 24 hours (S24). At the end of incubation period the samples are rinsed with multiple cycles of ethanol and DI water to remove any unbound thiols and finally blown dry with nitrogen. At this stage 50 $\mu$l of 80 nm diameter, citrate capped AuNS from BBI Solutions are dropcasted on the sample surface. The AuNS droplet on the functionalized Au mirror is removed after 30 seconds by blowing with nitrogen.

For STM imaging no AuNS are deposited on the samples to avoid unnecessary risk to the STM tip. Moreover, no markers are made on the mirror surface and therefore an unpatterned Si wafer is used for evaporation of Au thin film. The template stripping and the thiol functionalization procedure are the same as those mentioned above.

## Electrochemical (EC) characterization

EC characterization was performed using a single-compartment, gas-tight EC cell (BM EC model, redox.me). Before EC experiments, the cell was cleaned by Caro's acid and boiled three times in Milli-Q water. A silver/silver chloride electrode immersed in 3 M KCl (Ag/AgCl (3 M KCl), redox.me) was used as the reference electrode, a coiled Au wire was used as a counter electrode. All potentials in the text are referred to the scale versus Ag/AgCl (3 M KCl) reference electrode. Freshly prepared BPhT SAMs on glass/Cr/Au substrates were used as working electrodes. Before each experiment, the electrolyte (0.1 M NaOH, 8 mL) and the assembled, gas-tight cell were purged with nitrogen for at least 20 minutes. Gas inlet was then lifted to the gas compartment above the electrolyte. Nitrogen was purged in the gas compartment during EC characterization experiments. Electrochemical measurements were performed using SP-50 (BioLogic) potentiostat. Thiol reductive electrodesorption was performed by scanning the potentials from open circuit potential to -1.2 V versus Ag/AgCl (3 M KCl) at a scan speed of 50 mV/s in 0.1 M nitrogen-saturated NaOH.

## Scanning Tunneling Microscopy

Scanning tunneling microscopy (STM) experiments were performed using MS10 STM (Bruker) coupled with NanoScope Controller V (Bruker) and NanoScope software version 8.15 in constant current mode. Before each experiment, the sample stage and the top contact were thoroughly cleaned using ethanol and then, dried with nitrogen. Tips were prepared upon mechanically cutting Pt/Ir wire (0.25 mm in diameter, GoodFellow). To avoid any contamination, a new tip was prepared for each experiment. STM characterization of both S2 and S24 samples was performed on freshly prepared samples. The STM experiment was started within two hours of the sample preparation. BPhT SAM sample prepared upon 24 hr incubation (S24) was first characterized in air after which 2 $\mu L$ of nonanoic acid was added to the center of the sample surface to facilitate high-resolution STM imaging. We show both STM images obtained at the air/solid and at the nonanoic acid/solid interface. No significant differences in SAM structures were observed after adding nonanoic acid. STM images were analyzed using WSxM software.[68]

## Optical setup

The experimental setup consists of a confocal microscope designed to perform Raman and DF scattering spectroscopy. The laser exposure is carried out by a continuous wave (CW) Titanium Sapphire (Ti:Sapph) laser from Spectra Physics (Model 3900, pumped by Millenia 5W) tuned at 710 nm. The laser line is filtered using a bandpass filter (BP) (Fig. S6). Afterwards a lambda-half plate (LH), a polarizing beam-splitter (PBS) and a sliding neutral density filter (ND) are used to control the power of the laser beam. The LH and the PBS allow to decrease the laser power in large steps while the ND enables finer tuning. A pellicle beam-splitter (P) then reflects about 10% of the incident laser power towards a high NA objective (Olympus MPLFLN 100x 0.9NA) used to focus the laser on the NPoM and collect the Raman signal from the nanocavity-coupled molecules. The laser beam transmitted through the BS falls on a power meter (PM), thus the laser power can be monitored during exposure. The DF measurements are performed using a supercontinuum (NKT Photonics SuperK Compact) white light source focused on the sample by a secondary objective (Olympus Plan N 4x 0.1NA) at angle of 81 degree from the normal to the sample plan. In order to use the supercontinuum as an interference free white light source, its transverse spatial coherence length must be reduced. This is achieved by a procedure described elsewhere.[69] For all the measurements, signal from the sample is collected by the high NA objective. The acquired signal is split and a small portion is used to image the sample plane while the rest is passed through spectral filters and finally recorded using a spectrograph (Andor Kymera 193i) with CCD (Andor iDus). In case of Raman measurements, to remove laser reflection and Rayleigh scattering, the acquired signal is split by 90T/10R beam-splitter (BS); where transmission and reflection are used to filter out anti-Stokes (short pass: SP) and Stokes (long pass: LP) signals, respectively. After the filters, the anti-Stokes and Stokes are first aligned to be vertically parallel using a D-shape mirror and afterwards focused onto the spectrometer slit. For the collection of scattering spectra, the SP filter is moved out of the beam path and the reflection from the BS is blocked by a shutter. The binning region on the CCD is also reduced to act as a virtual spatial filter and minimize the collection area on the sample.

In order to compensate for the spectral response of the system in DF measurement the sample was replaced by a silver mirror tilted at 40 degrees on the horizontal plane with respect to the optical axis of the high NA objective. The specular reflection of the white light source is then recorded with the spectrometer. The acquired spectra are averaged, smoothed and normalized to obtain the DF spectral response. During the DF measurement of NPoMs, after baseline subtraction the scattering spectra are divided by the DF spectral response. To estimate the spectral response for the Raman measurements, a calibrated white light source (SLS201L ThorLabs) was used to illuminate the system via the high NA objective. The LP and SP filters are removed and spectra are collected separately for Stokes and antiStokes paths, hence the spectral response is computed separately for each path. The acquired spectra are then averaged, smoothed and divided by the calibration spectrum. These are then normalized by the maximum of antiStokes side In order to obtain the normalized spectral response of Stoke and antiStokes path. In order to correct a Raman spectrum both response spectra are combined at the laser wavelength and then used to divide the Raman spectrum.

# Acknowledgement

The authors thank Huatian Hu for his help regarding BEM simulations.

# Supporting Information Available

BPhT SAM heterogeneity on S24: Co-existance of different BPhT SAM structural phases along the neighbouring Au crystalline facets, Au surface reconstruction on S24, Molecular ordering on the S24 reconstructed facets, Regions with densely packed SAM on S2, Characterization of the BPhT SAM molecular ordering in S2 and S24 based on electrochemical and STM characterization, Schematics of optical setup, Change in DF with constant laser power, Complete power dependent measurements: S24, Complete power dependent measurements: S2, Temperature extraction by anti-Stokes background, Raman spectra S2 and S24, Evolution of higher order dipole mode.

# Supporting Information

# SI 1: BPhT SAM heterogeneity on S24: Co-existance of different BPhT SAM structural phases along the neighbouring Au crystalline facets

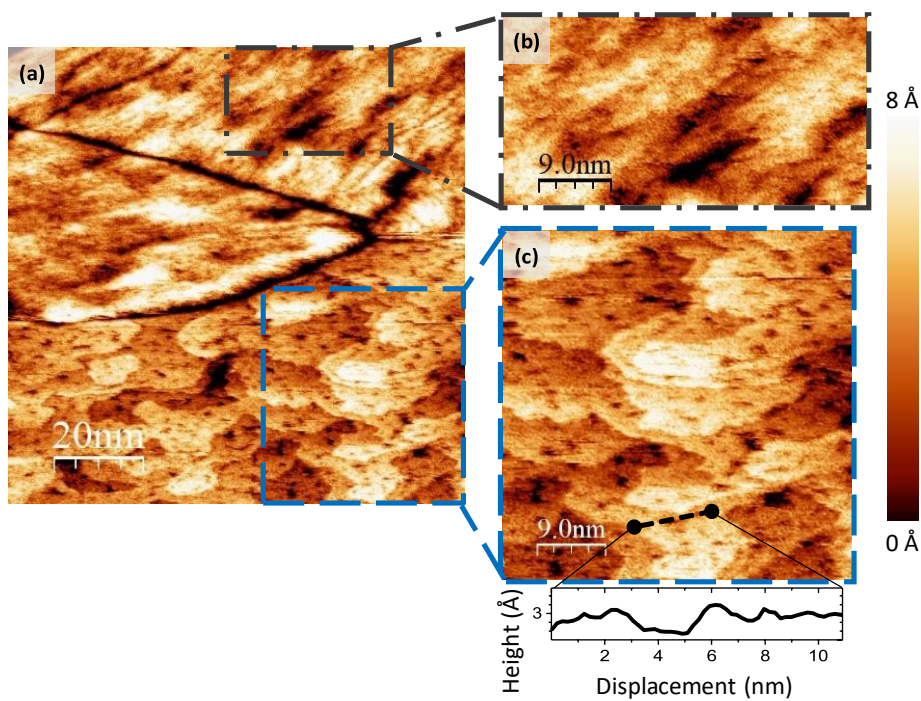

Figure S1: STM characterization of the BPhT SAM molecular ordering on S24. (a) Large-scale STM image showing different BPhT phases on three adjacent Au facets. (b) Zoom-in STM image of the upper facet shown in (a). No periodic arrangement, visible on STM images as the regular patterns, suggests that the degree of the molecular ordering is lower than for phase shown in (c).  (c) Zoom-in STM image of the lower facet shown in (a). The presence of the curved terraces (lateral size > 10 nm) and dark, Au vacancy islands (pits) randomly distributed along them are morphological features characteristic for the surface reconstruction that proceeds parallel to the SAM re-organization.[51] All images were obtained at the air/solid interface. Scanning parameters: $V_{bias}$=-200 mV, $I$ =338.2 pA

STM images in Figure S1(a) show significant differences between the two upper facets and the lower facet. Two upper facets appear nanostructured due to a few angstrom-large pro-trusions; yet, no periodic arrangement is observed. The lower facet appears nanostructured due to the curved terraces (lateral size > 10 nm) and dark, round features (i.e., pits formed upon reconstruction). These differences in the appearance of the crystalline facets imply that different SAM phases co-exist along S24. Heterogeneities in BPhT SAM ordering on the samples prepared upon prolonged incubation in ethanoic solutions is in agreement with earlier reports.[50,51] We speculate that BPhT SAMs on adjacent Au facets are in densely-packed phases with the small tilt angles between the principal molecular axis and the surface normal, as expected for SAMs prepared upon long immersion.[51] Also, we take this opportunity to remind the reader that the template stripped Au (i.e., the mirror in NPoM) is a polycrystalline surface with mostly (111) crystalline facets - it contains (100) and (110) facets and it is not atomically flat. Each of these facets might have a different SAM structure (in term of the molecular configuration); domains with weakly adsorbed double layers or even multilayer structures are also possible, as recently reported for similar system.[47] The direct consequence of the heterogeneities in S24 BPhT SAM ordering is evident in DF measurement in Figure 1(f) and 2(c, d) in the main text. Each NPoM on S24, prepared upon drop-casting NPs on the S24 BPhT SAM-functionalized mirror (shown here), has a different gap height, as evident from the large deviations in the peak position of the dipole mode.

## SI 2: Au surface reconstruction on S24

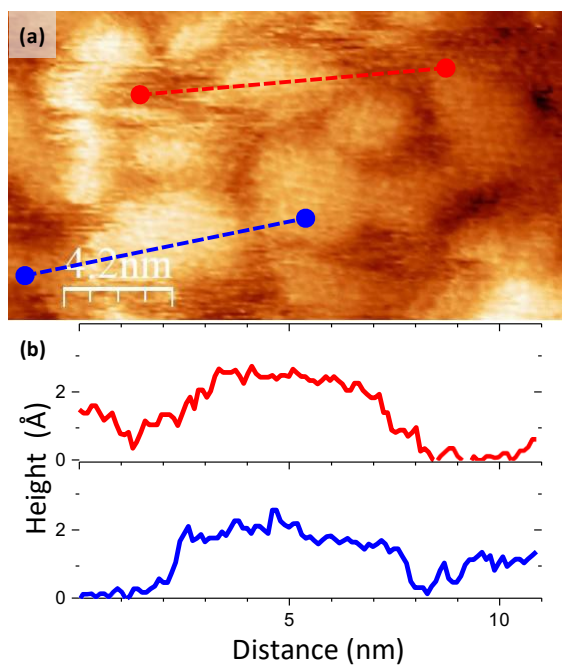

Figure S2: (a) STM image of S24 showing Au islands formed upon surface reconstruction. The apparent height and width of islands are $\simeq$ 2.4 Å and $\simeq$ 5 nm respectively, as shown in height profiles in (b). Note that the herringbone pattern of the BPhT SAM is visible on the islands' tops: this suggests that these islands are formed upon motion of the thiolate-Au adatom moieties. The formation of these islands is closely related to the formation of the vacancy depressions shown in Figure S1. This image was obtained at the nonanoic acid/solid interface. Scanning parameters: $V_{bias}$=-200 mV, $I$ =338.2 pA.

# SI 3: Molecular ordering on the S24 reconstructed facets

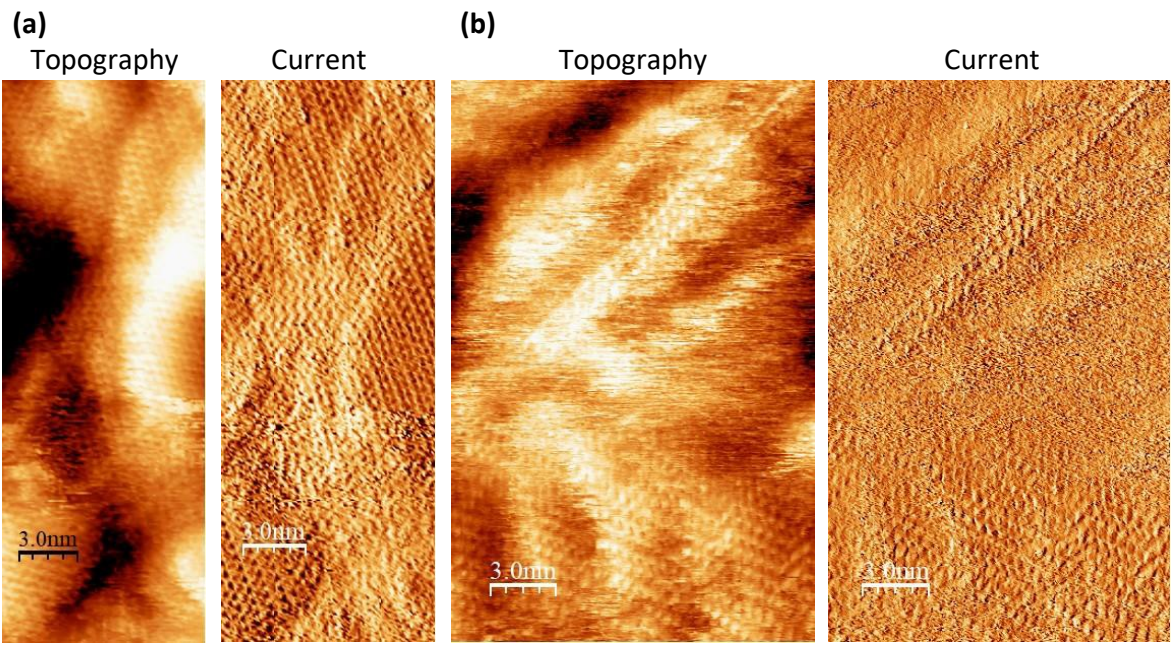

**(a)** Topography    Current    **(b)** Topography    Current

Figure S3: (a) STM topography and current images showing the same Au terraces on S24. Note the herringbone pattern extending over the terraces. This suggests that the SAM domains are not confined by the step edges of the terraces (or the small islands shown in Figure S2). Single SAM domain extends over, at least, 10 nm; however, multiple SAM domains are present along the single Au crystalline domain regardless of its orientation or lateral size. (b) STM topography and current images showing two domains rotated by $\approx$ 120°. Both herringbone pattern and rotational domains under $\approx$ 120° are characteristic for $(2\sqrt{3}\text{x}\sqrt{3})$.

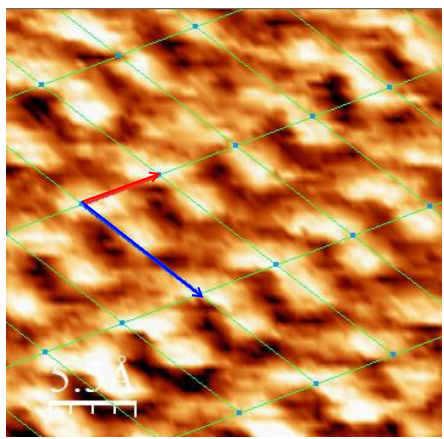

Figure S4: Herringbone pattern of BPhT SAM on S24 together with the superimposed lattice and the $(2\sqrt{3}\text{x}\sqrt{3})$ unit cell.

# SI 4: Regions with densely packed SAM on S2

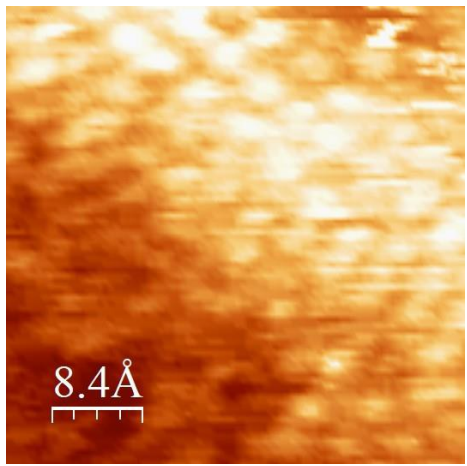

Figure S5: STM topography image showing the domains with densely packed BPhT SAM on S2. Scanning parameters: $V_{bias}$=-573 mV, $I$=494 pA.

# SI 5: Characterization of the BPhT SAM molecular ordering in S2 and S24 based on electrochemical and STM characterization

Table S1: Position, full-width at half-maximum ($FWHM$ ), and charge density ($Q$ ) of the peaks in the voltammograms of Figure 1(a) in the main text

| Properties | S2 | S24 |
|---|---|---|
| Potential versus Ag/AgCl (V) | -0.85 | -0.88 |
| $FWHM$ (V) | 0.07 | 0.08 |
| $Q$ ($\mu$C/cm$^2$) | 65 | 72 |

The parameters for the reductive desorption peaks shown in Figure 1(a) in the main text are listed in Table S1. The main differences in the integrated charge density − being 65.2 $\mu$C/cm$^2$ for S2 and 72.4 $\mu$C/cm$^2$ for S24 − imply on the less dense BPhT SAM layer on S2 comparing to the BPhT SAM layer on S24, further supporting discussion in the main text. Even though these estimates on integrated charge are not directly comparable to the literature on electrochemical characterization of aromatic thiol SAMs[70,71] because the glass/Cr/Au (mirror) used in this work were not flame annealed, formation of the less dense phase on S2 is in agreement with literature on characterization of similar systems prepared upon immersion in thiol solutions[50,51] or evaporation.[52]

Interpretation of reductive desorption, disclosing differences in the BPhT SAM ordering on global scale, is not straight forward due to the fact that several BPhT SAM phases co-exist on the neighbouring crystalline facets (see section SI 1). BPhT SAM heterogeneity is especially important for interpretation of the optical measurements on single NPoM revealing large deviations in the $l_1$ peak, mostly likely due to the differences in SAM (gap) heights for each NPoM. We thus evaluate the molecular ordering on S24 and S2 from STM images revealing the presence of the highly ordered BPhT SAMs across several hundred nanometers over the single crystalline facet on S24 and confined to the domains with the lateral dimensions < 10 nm on S2 together with disordered BPhT SAM on S2.

The distances between protrusions on S24, as-measured from the height profiles superimposed on the molecular rows in Figure 1 in the main text, are $8.9 \pm 0.9$ Å along the single row and $4.2 \pm 1.0$ Å across two rows implying on the square lattice close to the $(2\sqrt{3} \times \sqrt{3})$ structure (Figure S4). This finding is in excellent agreement with the paper by Azzam *et al.*[51] and Leung *et al.*[50] reporting on formation of SAMs with $(2\sqrt{3} \times \sqrt{3})$ structure and small domains upon long immersion in ethanolic solutions at room temperature. The phase with $(2\sqrt{3} \times \sqrt{3})$ structure, denoted as phase by Azzam *et al.*,[51] is characterized by a tilt angle between the molecular axis and the surface normal < 20° [50,72] and islands formed upon surface reconstruction (Figure S2).

Highly ordered BPhT SAMs domains on S2, having lateral size < 10 nm, co-exists with disordered phase shown in Figure 1 (c) in the main text. Interestingly, poor resolution of the disordered phase is in full agreement with references[50,51] reporting on poor quality of STM images of the BPhT adlayers.

# SI 6: Optical setup

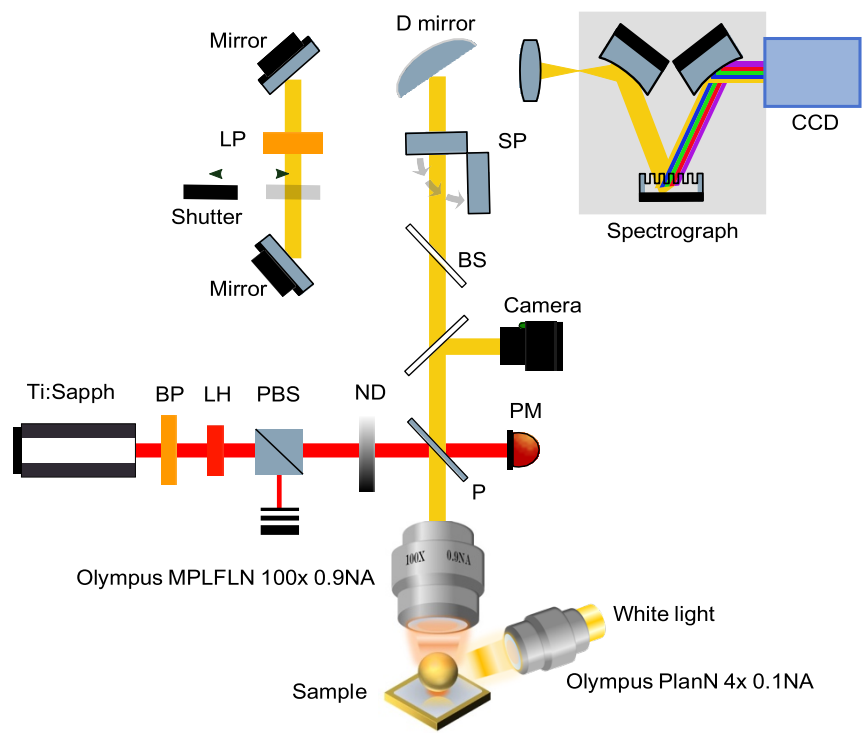

Figure S6: Dark field (DF) and Raman setup used for optical characterization.

# SI 7: Change in dark field (DF) with constant laser power

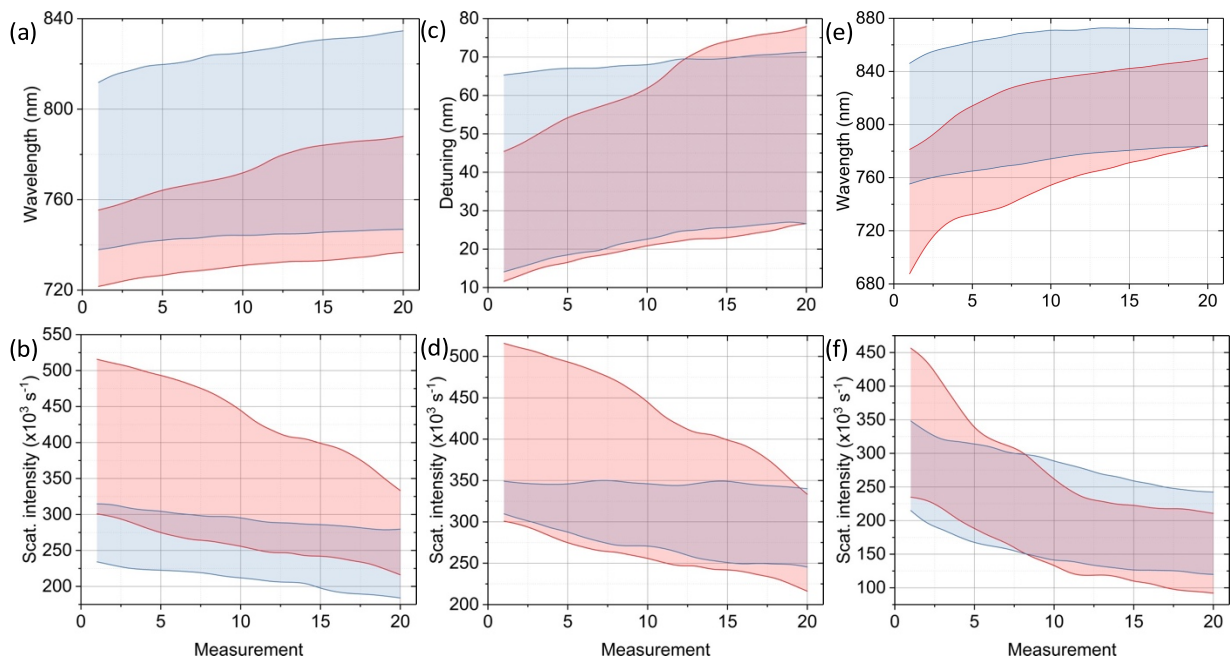

Figure S7: Peak wavelength and peak intensity of the dipole mode as a function of multiple exposures at constant laser power. The duration of exposure was limited to 5 minutes for each measurement. The red (blue) shaded region represents the standard deviation centered at the mean extracted by analyzing multiple NPoMs prepared with 24hr (2hr) incubation. (a,b) Constant power exposure with laser wavelength at 710 nm. (c,d) Constant power exposure with laser wavelength at 760 nm. Please note that the y-axis in (c) calculated by subtracting the laser wavelength from the peak position of the dipole mode. (e,f) Constant power exposure with laser wavelength at 632 nm. In all the experiments S24 NPoMs shows a red shift of the dipole peak along with a significant decrease in peak height. On the other hand S2 NPoMs show meagre changes in the position and peak height of the dipole mode. During the exposure at 632 nm both S2 and S24 show a much larger change in the peak position and intensity than observed for exposure with 710 and 760 nm. This could be the result of increased absorption of AuNS leading to significantly large temperature within the gap and hence prompting a faster molecular reorientation in S2 and S24.

# SI 8: Complete power-dependent measurements: S24

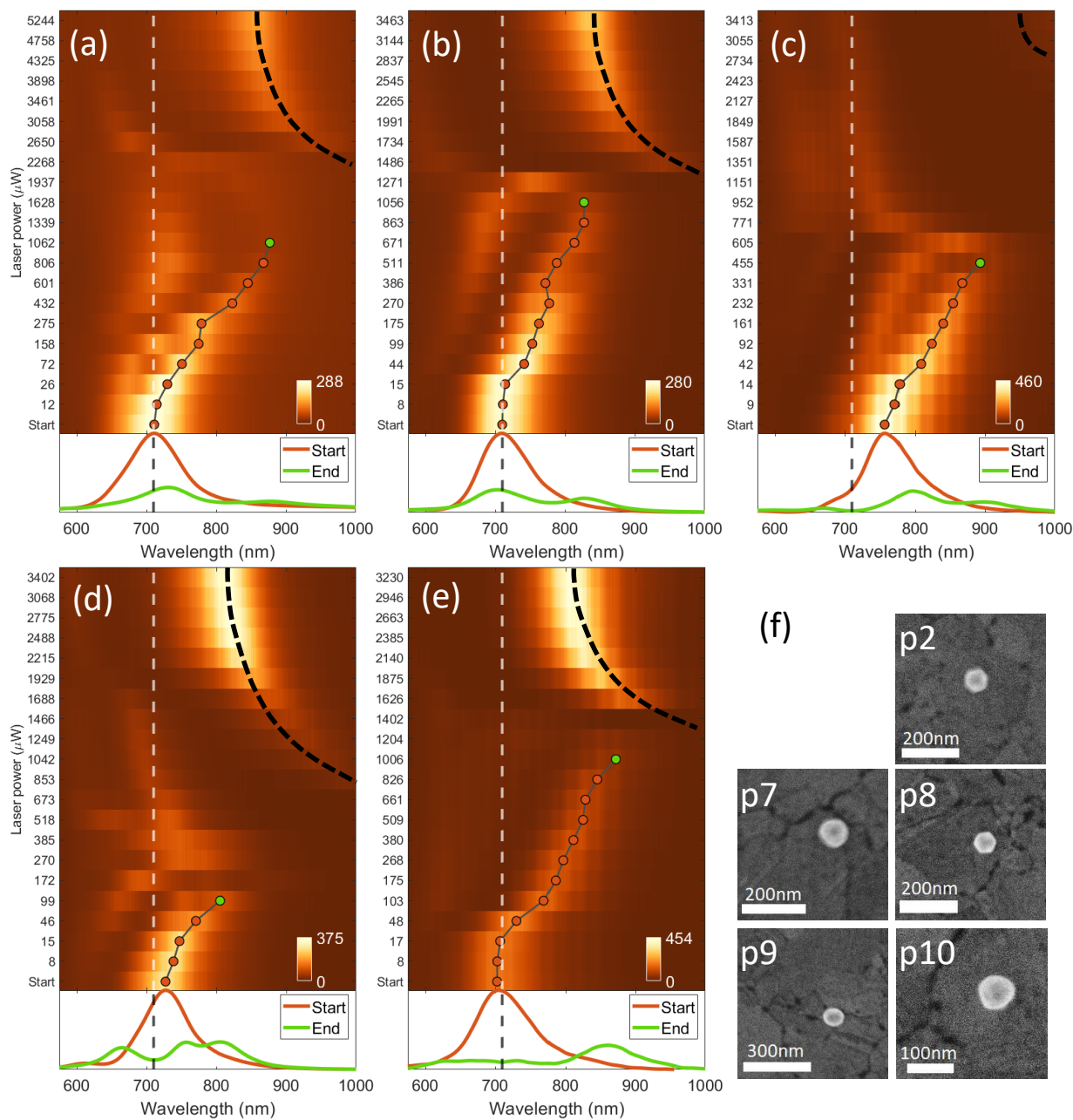

Figure S8: (a-e) Complete evolution of the DF spectra as the power of the exposure laser is increased. The spectral position of the dipole mode is shown by dot-dash line until the modes can be distinctly identified (green dot). The spectra at the first (red line) and last dot (green line) are shown in the figure below each contour plot. The fusion of the AuNP and the gold substrate underneath at high laser intensities can also be seen by the emergence on charge transfer plasmon (CTP) indicated by black dashed line. The plots (a-e) correspond to NPoMs consist of single AuNPs (p2-p10), as shown by the SEM images in (f), placed on high incubation time SAM.

# SI 9: Complete power-dependent measurements: S2

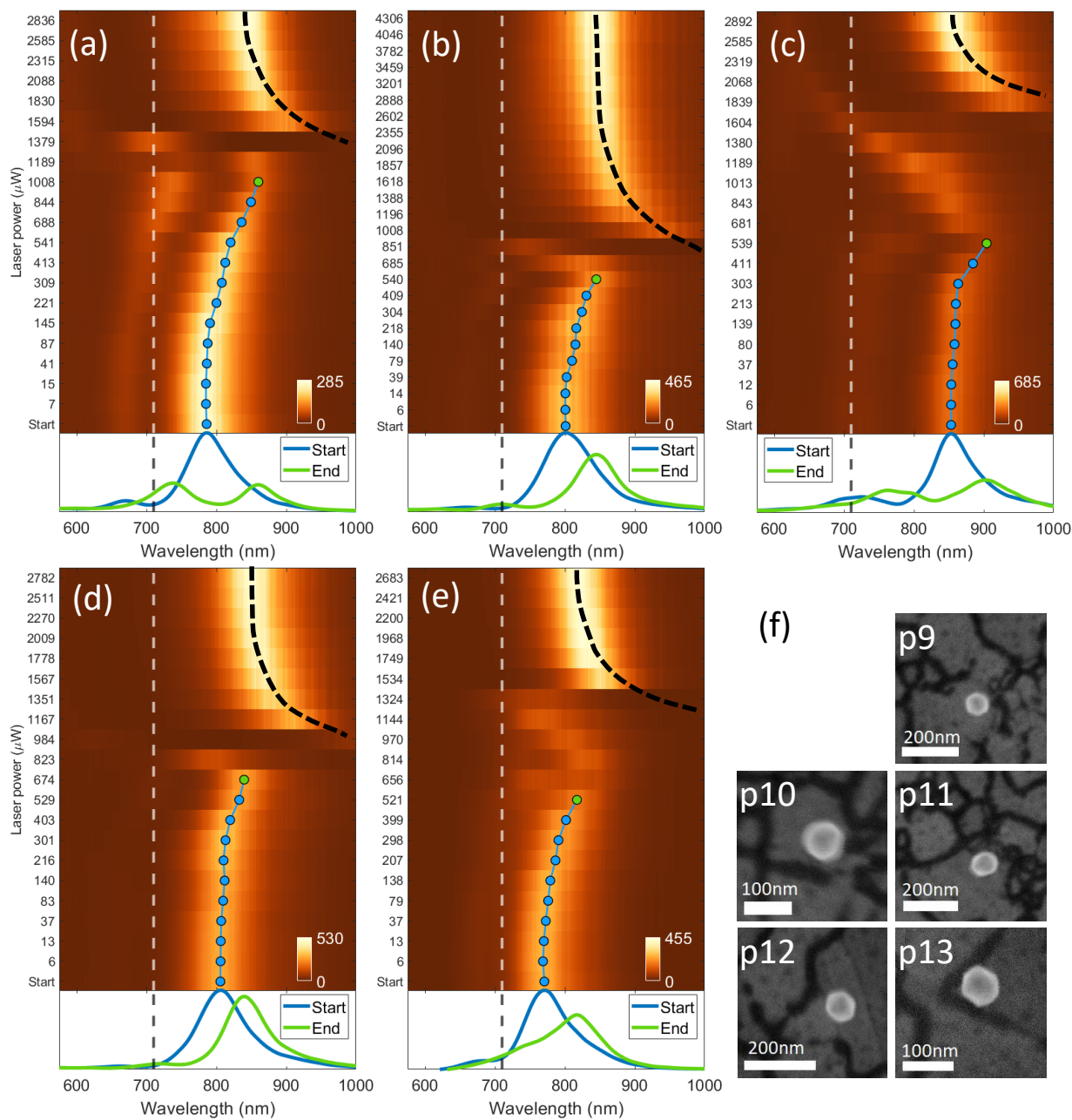

Figure S9: (a-e) Complete evolution of the DF spectra as the power of the exposure laser is increased. The spectral position of the dipole mode is shown by dot-dash line until the modes can be distinctly identified (green dot). The spectra at the first (blue line) and last dot (green line) are shown in the figure below each contour plot. The fusion of the AuNP and the gold substrate underneath at high laser intensities can also be seen by the emergence on charge transfer plasmon (CTP) indicated by black dashed line. The plots (a-e) correspond to NPoMs consist of single AuNPs (p9-p13), as shown by the SEM images in (f), placed on low incubation time SAM.

# SI 10: Temperature extraction by anti-Stokes background

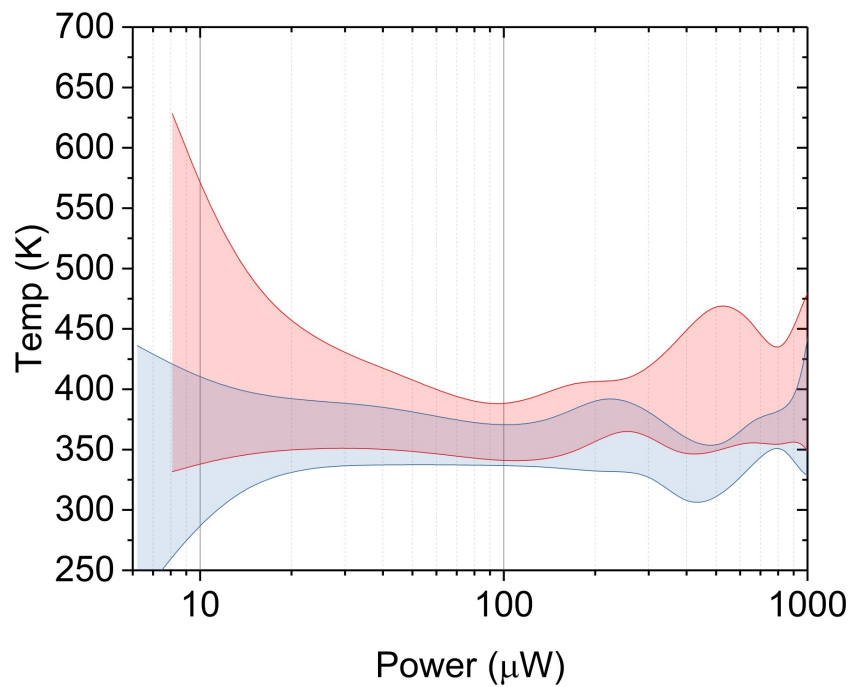

Figure S10: Electronic Raman temperature calculated by fitting the anti-Stokes background of multiple NPoMs with two exponentials. The large variations in temperature at power below 10 $\mu$W can be ignored due to poor signal to noise ratio. The initial temperature for NPoMs with low incubation time SAM was found to lower than those prepared with higher incubation time. Hence, the heat present within the nanogap could explain the molecular re-orientation and decrease in gap height.

# SI 11: Raman spectra S2 and S24

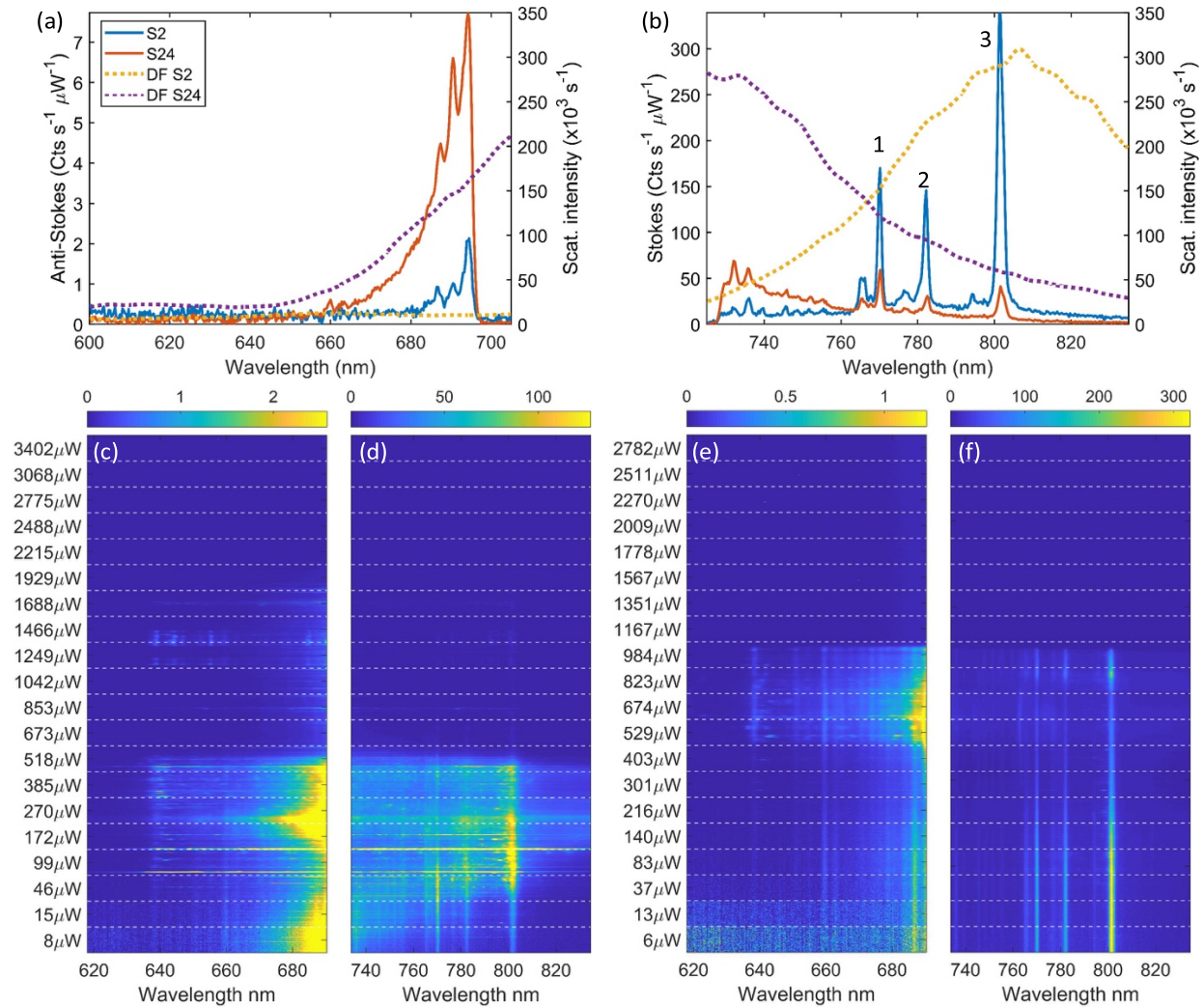

Figure S11: Anti-Stokes (a) and Stokes (b) spectra acquired on S2 (blue) and S24 (red) with laser wavelength at 710 nm and power 6 $\mu$W and 8 $\mu$W respectively. The signature Raman peaks of BPhT (1: 1079 $cm^{-1}$, 2: 1281 $cm^{-1}$, 3: 1586 $cm^{-1}$) are clearly visible in the Stokes sideband. The corresponding DF spectra for S2 (yellow) and S24 (purple) are also overlaid as dotted lines. The Raman signal is greatly influenced by the position of the plasmonic resonance. (c-f) Time series of anti-Stokes and Stokes spectra collected at different laser powers. The dashed white lines represent the change between laser powers. The color scale represents the signal recorded on the CCD as counts $s^{-1}$ $\mu W^{-1}$. (c-d) show anti-Stokes and Stokes corresponding to the S24 DF spectra in Figure 2(a) in the main text. (e-f) show anti-Stokes and Stokes corresponding to the S2 DF spectra in Figure 2(b) in the main text.

# SI 12: Evolution of higher order dipole mode

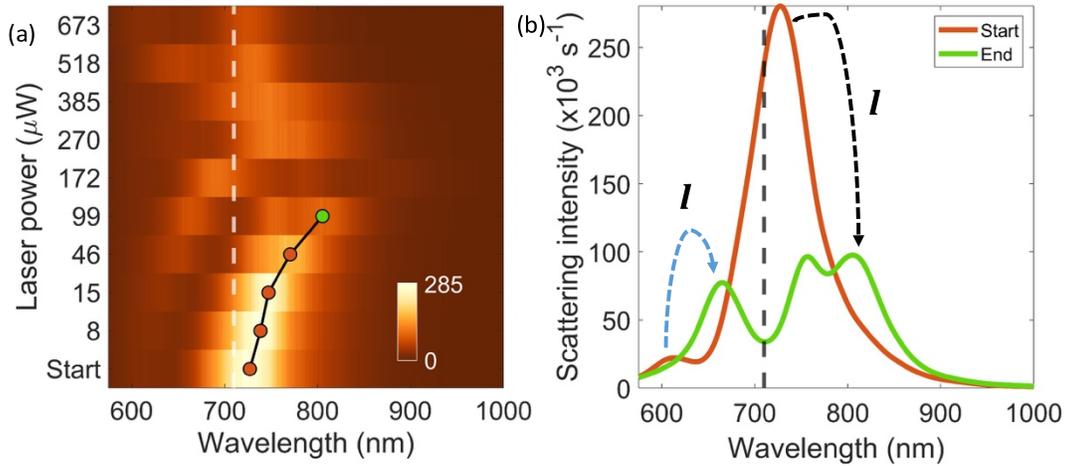

Figure S12: (a) The evolution of DF spectra on S24 as the laser power is increased. The red dot indicate the evolution of the bonding dipolar mode $l_1$ however after 99$\mu$W is it not possible to identify $l_1$. (b) Comparing the first (red) and the last (green) DF spectra show a redshift and decrease in the peak intensity of $l_1$. The higher order dipole mode $l_2$ also redshifts but shows an increase in its intensity marked by blue dashed arrow.